\documentclass[12pt]{article}
\usepackage{amssymb}

\makeatletter

\newcommand{\eqref}[1]{(\ref{#1})}
\newcommand{\Secref}[1]{\expandafter\MakeUppercase\secrefname~\ref{#1}}
\newcommand{\Appref}[1]{\expandafter\MakeUppercase\appendixname~\ref{#1}}
\newcommand{\Figref}[1]{\expandafter\MakeUppercase\figurename~\ref{#1}}
\newcommand{\Tbref}[1]{\expandafter\MakeUppercase\tablename~\ref{#1}}
   \let\secref\Secref  \let\appref\Appref
  \let\tbref\Tbref
\newcommand{\secrefname}{Section}
\DeclareSymbolFont{AMSb}{U}{msb}{m}{n}
\DeclareSymbolFontAlphabet{\mathbb}{AMSb}
\DeclareMathAlphabet\mathfrak{U}{euf}{m}{n}
\SetMathAlphabet\mathfrak{bold}{U}{euf}{b}{n}
\DeclareSymbolFont{boldletters}  {OML}{cmm} {b}{it}
\DeclareSymbolFont{boldoperators}{OT1}{cmr}{bx}{n}
\DeclareSymbolFont{boldsymbols}  {OMS}{cmsy}{b}{n}
\newcommand{\inv}{^{\raise.15ex\hbox{${\scriptscriptstyle -}$}\kern-.05em 1}} 
\newcommand{\ext}[1][]{\mathop{\raisebox{.2ex}{$\textstyle\bigwedge$}^{#1}}}  
\newcommand{\del}{\partial}
\newcommand{\dint}[2][]{\mathop{\mathalpha{\int#1}#2}}    
\newcommand{\lb}{\mathopen{[\![}}
\newcommand{\rb}{\mathopen{]\!]}}
\newcommand{\set}[1]{\mathbb{#1}}                         
\newcommand{\R}{\set{R}}
\def\C{\set{C}}

\newcommand{\group}[1]{\mathop{\kern\z@\mathrm{#1}}\nolimits}     

\newcommand{\GL}{\group{GL}}
\newcommand{\lie}[1]{\mathop{\kern\z@\mathfrak{#1}}\nolimits}
\newcommand{\g}{{\lie{g}}}
\newcommand{\h}{{\lie{h}}}
\renewcommand{\u}{{\lie{u}}}

\newcommand{\opname}[1]{\mathop{\kern\z@\mathrm{#1}}\nolimits}    
\newcommand{\Tr}{\opname{Tr}}
\newcommand{\Hom}{\opname{Hom}}
\newcommand{\End}{\opname{End}}
\newcommand{\Aut}{\opname{Aut}}

\newcommand{\eps}{\epsilon}
\newcommand{\CB}{\mathcal{B}}   
\newcommand{\CF}{\mathcal{F}}   
\newcommand{\CH}{\mathcal{H}}   
\newcommand{\CK}{\mathcal{K}}   
\newcommand{\CL}{\mathcal{L}}   
\newcommand{\CN}{\mathcal{N}}   
\newcommand{\CP}{\mathcal{P}}   
\newcommand{\CX}{\mathcal{X}}   
\newcommand{\fA}{\mathfrak{A}}
\newcommand{\fB}{\mathfrak{B}}
\newcommand{\fC}{\mathfrak{C}}
\newcommand{\fH}{\mathfrak{H}}
\DeclareMathSymbol{\Bd}{\mathalpha}{boldletters}{`d}
\DeclareMathSymbol{\BA}{\mathalpha}{boldletters}{`A}
\DeclareMathSymbol{\BB}{\mathalpha}{boldletters}{`B}
\DeclareMathSymbol{\BF}{\mathalpha}{boldletters}{`F}
\DeclareMathSymbol{\BH}{\mathalpha}{boldletters}{`H}
\DeclareMathSymbol{\BK}{\mathalpha}{boldletters}{`K}
\DeclareMathSymbol{\BM}{\mathalpha}{boldletters}{`M}
\DeclareMathSymbol{\BN}{\mathalpha}{boldletters}{`N}
\DeclareMathSymbol{\BQ}{\mathalpha}{boldletters}{`Q}
\DeclareMathSymbol{\BU}{\mathalpha}{boldletters}{`U}
\DeclareMathSymbol{\BV}{\mathalpha}{boldletters}{`V}
\DeclareMathSymbol{\BW}{\mathalpha}{boldletters}{`W}
\DeclareMathSymbol{\BX}{\mathalpha}{boldletters}{`X}
\DeclareMathSymbol{\Bphi}{\mathord}{boldletters}{"1E}
\renewcommand{\section}{\@startsection{section}{1}{\z@}%
                                    {-7ex \@plus -1ex \@minus -.2ex}%
                                    {2.5ex \@plus.2ex}%
                                    {\normalfont\large\scshape\centering}}
\renewcommand{\subsection}{\@startsection{subsection}{2}{\z@}%
                                       {-5ex \@plus -1ex \@minus -.2ex}%
                                       {1.5ex \@plus.2ex}%
                                       {\normalfont\normalsize\scshape}}
\newcommand{\sectionname}{}

\renewcommand{\appendix}{\par
  \setcounter{section}{0}%
  \setcounter{subsection}{0}%
  \renewcommand{\thesection}{\@Alph\c@section}%
  \renewcommand{\sectionname}{\appendixname}}
\renewcommand\@seccntformat[1]{\ignorespaces\csname #1name\endcsname\space
                               \csname the#1\endcsname.\quad}   
\newdimen\captionmargin 
\newdimen\captionindent 
\newdimen\captionwidth 
\newcommand{\captionfont}{\slshape}
\newcommand\@captionlabel[1]{\textsc{#1:}\space}
\long\def\@makecaption#1#2{%
  \vskip\abovecaptionskip  
  \captionwidth\hsize 
  \advance\captionwidth -2\captionmargin
  \sbox\@tempboxa{\@captionlabel{#1}\captionfont #2}%
  \ifdim \wd\@tempboxa >\captionwidth
    \ifdim\captionindent>\z@ 
      \advance\captionwidth -\captionindent
      \hskip\captionindent
    \fi
    \hskip\captionmargin
    \parbox[t]{\captionwidth}{\leavevmode\hskip-\captionindent
      \@captionlabel{#1}\captionfont #2}%
  \else
    \global \@minipagefalse
    \hb@xt@\hsize{\hfil\box\@tempboxa\hfil}%
  \fi
  \vskip\belowcaptionskip}
\def\eqnarray{%
   \stepcounter{equation}%
   \def\@currentlabel{\p@equation\theequation}%
   \global\@eqnswtrue
   \m@th
   \global\@eqcnt\z@
   \tabskip\@centering
   \let\\\@eqncr
   $$\everycr{}\halign to\displaywidth\bgroup
       \hskip\@centering$\displaystyle\tabskip\z@skip{##}$\@eqnsel
      &\global\@eqcnt\@ne$\;\hfil{##}$\hfil
      &\global\@eqcnt\tw@$\;\displaystyle{##}$\hfil\tabskip\@centering
      &\global\@eqcnt\thr@@ \hb@xt@\z@\bgroup\hss##\egroup
         \tabskip\z@skip
      \cr
}
\ifcase \@ptsize
\or 
\or 
\fi
  \divide\@tempdima\baselineskip
  \@tempcnta=\@tempdima
\makeatother

\begin{document}

%
%

\thispagestyle{empty}

\begin{flushright}\scshape
RUNHETC-2002-24\\
hep-th/0207017\\
June 2002
\end{flushright}
\vskip1cm

\begin{center}

{\LARGE\scshape
Nonabelian 2-Forms
\par}
\vskip15mm

\textsc{Christiaan Hofman}
\par\bigskip
{\itshape
  New High Energy Theory Center, Rutgers University,\\
  136 Frelinghuysen Road, Piscataway, NJ 08854, USA,}
\par\bigskip
\texttt{hofman@physics.rutgers.edu}

\end{center}

\vspace*{25mm}

\section*{Abstract}

We study gauge theories based on nonabelian 2-forms. 
Certain connections on loop space give rise to generalized 
covariant derivatives that include a nonabelian 2-form. 
This can be used to find rather straightforward 
expressions for the field strength and gauge transformations. 
As a special case we recover formulas for connections 
on nonabelian gerbe, as recently constructed in mathematics. 
The general construction gives rise to connections on algebra 
bundles, which might be relevant for D-branes 
in the presence of torsion. We construct BV sigma model actions 
for these connections and discuss their gauge fixing. 
We find both Yang-Mills type and topological BF type actions.

\newpage
\setcounter{page}{1}
%
%

\section{Introduction}

Strings, being extended objects of dimensions two, 
naturally couple to 2-form connections. Indeed superstrings 
are always charged under a 2-form called the $B$-field. 
This 2-form is described at low energies by a Maxwell type 
gauge theory, which easily extends to higher $p$-forms such 
as the $B$-field. In contrast the nonabelian gauge theory 
seems almost impossible to extend to higher $p$-forms. 

Nonabelian 1-form gauge theories appear naturally in string theory 
on the worldvolume of a stack of D-branes. The nonabelian nature is 
related to the internal Chan-Paton indices labeling the branes. 
Two-forms also appear in the low energy description of extended object 
in string theory, namely NS 5-branes in Type IIA and M5-branes in M-theory. 
It has long been speculated that stacks of these 5-branes at low energy are described by 
nonabelian 2-forms. The decoupled low energy theories describing these 
systems in a particular limit are actually string theories, the $(2,0)$ 
little string theories (LST) \cite{mm}, see also \cite{aha} and references 
therein. They should exist for any gauge group in the A-D-E series. 
A strong hint for the appearance of nonabelian 2-forms in these 
systems comes from string duality. According to this, the dimensional 
reduction of the $(2,0)$ LST should give rise to $N=4$ super Yang-Mills. 
In the abelian case it can easily be seen how the (self-dual) 2-form 
reduces to the abelian 1-form gauge field. The question is how the 
nonabelian 1-form in the Yang-Mills lifts to a 2-form gauge field. 

In this paper we will try to find a consistent gauge theory for 
nonabelian 2-forms. The starting point will be gauge bundles 
on loop space. Connections on abelian bundles on loop space can be 
described in terms of abelian 2-forms---integrating a 2-form over 
the loop gives a 1-form on loop space. 
We will consider what happens if one tries to describe 
nonabelian bundles. We will find there is room for a nonabelian 2-form. 
Consistency in the gauge transformations immediately shows the need of 
an additional 1-form gauge field. The expressions that arise for the 
action of the covariant derivative can be understood in terms of the 
Hochschild complex of multilinear maps on the internal vector space. 
The connection form combines into a differential form of mixed degree 
with values in this complex. This complex is known to have the structure 
of a Lie algebra. This makes it surprisingly easy to write down expressions 
for curvature, gauge transformations, and the Bianchi identity. 

Abelian 2-forms are closely related to the theory of gerbes. It is 
therefore no surprise that gerbes have been shown to play a role 
in string theory \cite{sharpe1,sharpe2,kalk1,kalk2,frewit,seven,keur}. 
The conclusion is that the abelian 2-forms in 
string theory can indeed be understood as a connection on 
an abelian gerbe. Nonabelian generalizations of gerbes are much 
less understood. Global descriptions of nonabelian gerbes were 
already known for some time, a theory initiated by Gireaud \cite{gir}. 
Only recently mathematicians were able to define 
a notion of connections on these nonabelian gerbes \cite{bremes}. 
Their theory is based on 2-categories and gauge stacks. 
Similar approaches can also be found in \cite{attal,baez}, 
and a related global construction of nonabelian Wilson surfaces, 
inspired by this work, has appeared in \cite{chep}. 
In this paper we will take a somewhat different approach to the subject. 
Rather than using this rather abstract theory we will try to 
rely more on down to earth methods of differential forms. 
For a sound mathematical basis however 2-categories are probably 
needed, and in fact formed a motivation for this work. 

We will find a natural extension of 2-form connections 
to describe algebra bundles. In this case the product in the fiber 
can vary over the base space. Algebra bundles have appeared 
in recent years in string theory in the form of Azuyama algebras 
\cite{kapus,boumat}. These are bundles of matrix algebras $M_N(\C)$. 
They appeared in the context of twisted K-theory, which describes the charges 
of D-branes in the presence of a torsion $B$-field \cite{witk}. 
The effect can be understood by replacing the algebra of functions by 
the algebra of sections of the $M_N(\C)$ bundles. Hopefully the algebra 
bundles in our description can be used to describe more explicitly 
the role of these Azuyama algebras in string theory. 

Formalizing the general setup of algebra bundles we can try to reduce the 
gauge group to any group. There are however some restrictions, as one 
wants to keep the structure of differential Lie algebra that made 
it possible to write down a consistent theory. This will give rise 
naturally to so-called crossed modules. A crossed module is a pair of groups 
connected by a homomorphism, $G\stackrel\delta\to H$, and a compatible 
action of $H$ on $G$. The gauge theory will consist of a 2-form 
with values in the Lie algebra $\g$ and a 1-form with values in $\h$. 
This generalizes other approaches to nonabelian 2-forms 
\cite{alfegu,bau,lahi1,lahi2}, which were mostly related to the somewhat 
trivial crossed module with $H=G$. Crossed modules play an important 
role in the theory of nonabelian gerbes, so it is no surprise to find them here. 
In fact our formulas reduce to those of \cite{bremes} in this case. 
What is important to note is that in fact the data of nonabelian 2-forms 
is more than just the gauge group, one needs a crossed module instead. 
This extra data in fact does have some effect on possible fluxes 
that can arise in theories described by nonabelian 2-forms. These 
can be described topologically, and probably are the easiest 
signs of nonabelian 2-forms to detect in string theory. 

A next step we will undertake is to write down an action for the 
connection. Actually, the general Lie algebra structure makes it easy to 
write down a BV sigma model for the generalized connection. 
This model will however not directly give the physics of the gauge theory. 
For this we need to provide a gauge fixing fermion. This will give us 
several possible physical theories. One gauge fixing gives rise 
to a Yang-Mills type gauge theory. It is also possible 
to find topological interacting BF type terms, giving flat 
field strengths as their equations of motion. In 5 
dimensions one can also get a 2-form Chern-Simons type theory. 

This paper is organized as follows. In \secref{sec:loopforms} we 
give a description of differential geometry on loop space. This will 
motivate our main setup for describing generalized connections. 
In \secref{sec:algbundle} we use the result to describe the curvature, 
gauge tranformation, and Bianchi identity for algebra bundles. 
In \secref{sec:crosmod} we discuss reductions of the gauge group 
and the role of crossed modules. We will find that these reduced 
connections are closely related to the connections on gerbes 
found in \cite{bremes}. 
In \secref{sec:bv} we construct BV sigma models for the connections 
on algebra bundles and gerbes. We will conclude with some discussion 
and open problems.

\section{Forms and Bundles on Loop Space}
\label{sec:loopforms}

In this section we present a description of differential forms 
and bundles on loop space. This section mainly serves as a motivation 
for our nonabelian 2-form theory.

\subsection{Differential Forms on Loop Space}

Given a point $x_0\in X$, the (based) loop space is defined as the space 
of loops in $X$ starting at $x_0$, 
\begin{equation}
  \CL X_{x_0} = \{ x:S^1\to X \mid x(0)=x(2\pi)=x_0 \}.
\end{equation}
where we use the interval $[0,2\pi]$ to describe the circle. 
In the remainder we will mostly leave out the explicit reference to the 
base point $x_0$. It will be useful to view the values of the loop 
$x^\mu(\sigma)=x^\mu_\sigma$ as an infinite set of coordinates on loop space, 
viewing $\sigma$ as a label. Denoting by $\delta$ the De Rham differential on $\CL X$, 
this induces an infinite basis $\delta x^\mu(\sigma)=\delta x^\mu_\sigma$ 
for the 1-forms on loop space, induced by these coordinates. 

Our first objective is the description of differential forms on 
loop space.\footnote{We will consider only a restricted set of differential 
forms which vanish on constant loops, build out of forms that vanish on the 
base point $x_0$. More general differential forms can be described in 
a similar way, giving rise to a cyclic version \cite{getzjon1}.} 
The essential observation is that they can be constructed 
out of the differential forms on the space $X$ \cite{getjonpet}. 
Given a differential form $\omega$ on $X$ of degree $p$ we can construct a 
differential form on the loop space $\CL X$, which we denote $\oint(\omega)$, 
by pull back and integration, as 
\begin{equation}
  \oint(\omega) 
  = \dint[_{S^1}]{d\sigma} x_\sigma^*(\iota_{\dot x_\sigma}\omega)
  = \frac{1}{p!}\dint[_{S^1}]{d\sigma} \omega_{\mu_1\ldots\mu_{p-1}\mu_{p}}(x(\sigma))
  \delta x^{\mu_1}(\sigma)\cdots\delta x^{\mu_{p-1}}(\sigma)\dot x^{\mu_{p}}(\sigma),
\end{equation}
where $\dot x=\frac{dx}{d\sigma}$ and $\iota$ denotes right contraction. 
Note that this differential form has degree $p-1$, one lower than the original form. 
More generally, there is a construction which relates to a $n$-tuple of 
differential forms $\omega_1,\ldots,\omega_n$ a differential form 
by a generalization of this construction,  
\begin{equation}
  \oint(\omega_1,\ldots,\omega_n) \equiv 
  \dint[_{\Delta_n}]{d^n\!\sigma} x_{\sigma_1}^*(\iota_{\dot x_{\sigma_1}}\omega_1)\wedge\cdots\wedge x_{\sigma_n}^*(\iota_{\dot x_{\sigma_n}}\omega_n), 
\end{equation}
where we integrate over the $n$-simplex 
\begin{equation}
  \Delta_n = \{(\sigma_1,\ldots,\sigma_n)\in \R^n \mid 0\leq \sigma_1\leq\cdots\leq \sigma_n\leq 2\pi\}.
\end{equation}
The degree of this form on loop space is $\sum_{i=1}^k (p_i-1)$, 
where $p_i$ is the form degree of $\omega_i$. It can be shown that these generate 
the full algebra of differential forms on loop space (supplied with some 
completeness argument). This is intuitively clear from the fact that the 
$\delta x^\mu(\sigma)$ form a basis of the differential forms on loop space. 
Identifying the space of $n$-tuples of differential forms with elements 
of the tensor product $T(\Omega(X))=\bigoplus_n\Omega(X)^{\otimes n}$, 
we have constructed a map $\oint:T(\Omega(X))\to\Omega(\CL X)$. 

The natural operations on the algebra of differential forms on 
loop space, the De Rham differential and the wedge product, 
can be formulated in terms of the tensor algbera of forms using this map. 
The De Rham differential $\delta$ on loop space acts on the above 
forms as follows 
\begin{eqnarray}
  \delta\oint(\omega_1,\ldots,\omega_n) 
  &=& \sum_k(-1)^{\eps_{k-1}}\oint(\omega_1,\ldots,d\omega_k,\ldots,\omega_n)
\nonumber\\
  && + \sum_k(-1)^{\eps_{k}}\oint(\omega_1,\ldots,\omega_k\wedge\omega_{k+1},\ldots,\omega_n),
\end{eqnarray}
with $\epsilon_k=\sum_{i\leq k}(p_i-1)$. The terms involving the wedge 
product are a result of the relation 
$d(\iota\omega)=\iota( d\omega)-(-1)^p\frac{d}{d\sigma}\omega$. The integration 
of the last term gives boundary terms which can be written as wedge products. 
The other important operation is of course the wedge product. It is given by 
\begin{equation}
  \oint(\omega_1,\ldots,\omega_n)\wedge\oint(\omega_{n+1},\ldots,\omega_{n+m})
  = \sum_{\sigma\in S_{n,m}}(\pm)\oint(\omega_{\sigma\inv(1)},\ldots,\omega_{\sigma\inv(n+m)}),
\end{equation}
where the sum runs over all $(n,m)$-shuffles---permutations $\sigma\in S_{n+m}$ 
for which 
\begin{equation}
  \sigma(1)<\cdots<\sigma(n)\quad\mbox{and}\quad\sigma(n+1)<\cdots<\sigma(n+m).
\end{equation}
The sign is determined by the permutation and the degrees of the forms shifted by one. 
For example $\oint(\omega_1)\wedge\oint(\omega_2) = \oint(\omega_1,\omega_2)+(-1)^{(p_1-1)(p_2-1)}\oint(\omega_2,\omega_1)$.

\subsection{Gauge Bundles and Covariant Derivatives}

We will next discuss covariant derivatives on loop space. 
They will be described as a deformation of the De Rham differential. 
This can be accomplished by adding a 1-form space to $\delta$. 
Due to the shift in the degree by one a 2-form $B$ on $X$ 
induces a 1-form $\oint(B)$ on loop space, which we can 
then use to construct a covariant De Rham differential 
$\nabla=\delta+\oint(B)$. This is the standard way to 
write down a connection on line bundles over loop space. 
We can add more general 1-forms on loop space, 
involving $n$-tuples of 1-forms in addition to a 2-form. 
Below we will see that in the nonabelian case this will 
automatically occur. For now we make the simplifying 
ansatz above, involving only a single integral over the loop. 

The infinitesimal gauge transformations that are consistent 
with the above ansatz for the covariant derivative have the form 
\begin{equation}
  \delta_\alpha\oint(\omega_1,\ldots,\omega_n) = \oint(\alpha)\wedge\oint(\omega_1,\ldots,\omega_n)
  = \sum_k\oint(\omega_1,\ldots,\omega_k,\alpha,\omega_{k+1},\ldots,\omega_n),
\end{equation}
where $\alpha$ is a 1-form. The generators of the finite 
gauge transformations are found by exponentiation 
\begin{equation}
  \exp \oint(\alpha) = \sum_{n\geq0}\oint(\alpha,\stackrel{n}\ldots,\alpha)
  = \exp\Bigl(\dint[_0^{2\pi}]{d\sigma}\alpha_\mu(x(\sigma))\dot x^\mu(\sigma)\Bigr),
\end{equation}
where the $n$ above the dots denotes that $\alpha$ is repeated $n$ times. 
In other words, a finite gauge transformation is a Wilson line along the loop. 
The connection 2-form $B$ transforms under these gauge transformations 
in the expected way as $\delta_\alpha B=d\alpha$. 

For this paper we are ultimately interested in nonabelian generalizations 
of this construction. We want the connection 2-form $B$ taking values in some 
nonabelian algebra, for example, an algebra of matrices on some vector space. 
For the moment, let us not be specific, and just use some internal vector space $\g$. 
We will denote the combination of the product in $\g$ with the wedge products by $\star$. 
The action of $B$ by the wedge product on loop space shows that it gets 
inserted in the string of forms. Therefore, consistency requires us to 
take the $\omega_i$ also with values in $\g$. Note that the definition 
of the map $\oint$ involves taking the wedge product, which therefore 
has to be replaced by the nonabelian version $\star$. 

Let us check whether the gauge invariance is still consistent with our 
restricted choice of connection. The gauge variation of the connection $\nabla$ 
by a gauge transformation $\delta_\alpha$ can be calculated by the commutator 
of the gauge transformation with the covariant derivative. We find 
\begin{eqnarray}
  (\nabla\circ\delta_\alpha-\delta_\alpha\circ\nabla)\oint(\omega_1,\ldots,\omega_n)
  &=& \sum(-1)^{\eps_{k}}\oint(\omega_1,\ldots,\omega_k,d\alpha,\omega_{k+1}\ldots,\omega_n)\nonumber\\
&&  + \sum(-1)^{\eps_{k-1}}\oint(\omega_1,\ldots,i_\alpha\omega_k,\ldots,\omega_n),
\end{eqnarray}
where $i_\alpha$ denotes the supercommutator with $\alpha$, 
$i_\alpha\omega=\alpha\star\omega-(-1)^{p}\omega\star\alpha$. 
The first sum shows that the 2-form gauge field transforms as expected, 
$\delta_\alpha B=d\alpha$. However the terms involving $i_\alpha$, which vanish 
in the abelian case, can not be absorbed in a gauge transformation of $B$. 
Rather, it can be viewed as a change in the action of the differential 
$d$ to $d+i_\alpha$. This shows that we should also make $d$ covariant. 
The problem arises because our definition of $\oint$ was not covariant. 
Note that we multiply forms at different points along the loop. 
We assumed that the $\g$ was trivial over the manifold $X$, and hence 
also over the loop. Now we see that in the nonabelian case this assumption 
is inconsistent, and we need to consider $\g$ as a bundle rather than a 
fixed vector space. 

When we turn $\g$ into a bundle we need to connect the different points 
on the loop by a Wilson line. Let us choose a 1-form gauge field $A$ on $X$ 
and introduce the Wilson line along the loop 
\begin{equation}
  W_\sigma = \CP\exp\Bigl(\dint[_0^\sigma]{d\sigma'} A_\mu(x(\sigma'))\dot x^\mu(\sigma')\Bigr).
\end{equation}
Using these Wilson lines we define covariant pull backs of the forms $\omega$ as 
\begin{equation}
  \omega^W(\sigma) = W_\sigma\inv(x_\sigma^*(\iota_{\dot x_\sigma}\omega)).
\end{equation}
Note that the action of $W$ pulls back the form to the fiber at $x(0)$. 
Hence multiplication can be done covariantly and locally, using the product 
in this fiber. This leads to a covariant generalization of the map $\oint$ 
given by 
\begin{equation}
  \oint_A(\omega_1,\ldots,\omega_n) 
  = \dint[_{\Delta_n}]{d^n\!\sigma}\omega_1^W(\sigma_1)\star\cdots\star\omega_n^W(\sigma_n),
\end{equation}
where $\star$ is the product, build out of the product $M$ evaluated in 
the fiber at $x(0)$ and the wedge product. Note that for $n=1$ we have an expansion 
\begin{equation}
  \oint_A(\omega) = \sum_n(-1)^n\oint(A,\stackrel{n}\ldots,A,\omega).
\end{equation}
This covariant form should also be applied to the connection 2-form, 
i.e.\ we now have $\nabla=\delta+\oint_A(B)$.
Explicitly, the action of the covariant derivative becomes 
\begin{eqnarray}
  \nabla\oint_A(\omega_1,\ldots,\omega_n)
  &=& \sum(-1)^{\eps_{k-1}}\oint_A(\omega_1,\ldots,d\omega_k,\ldots,\omega_n)
\nonumber\\
  && +\sum(-1)^{\eps_{k}}\oint_A(\omega_1,\ldots,\omega_k\star\omega_{k+1},\ldots,\omega_n)
\nonumber\\
  && +\sum(-1)^{\eps_{k-1}}\oint_A(\omega_1,\ldots,A(\omega_k),\ldots,\omega_n)
\\
  && +\sum(-1)^{\eps_{k}}\oint_A(\omega_1,\ldots,\omega_k,B,\omega_{k+1},\ldots,\omega_n).\nonumber
\end{eqnarray}
Here $A(\omega)$ combines the wedge product and the action of the gauge 
field on the algebra $\g$. Notice that $A$ is valued in $\End(\g)$. 
We can now absorb the troublesome term in the gauge 
transformation as a gauge transformation of $A$ if we take 
$\delta_\alpha A = i_\alpha$. 

The explicit formula shows that we can formally write the covariant 
derivative as $\nabla=d+M+A+B$, where $M$ is the product. Each 
of the terms is interpreted as a multilinear map 
$T^n(\Omega(X,\g))\to \Omega(X,\g)$, acting on the full 
tensor algebra $T(\Omega(X,\g))$, with the action of a general 
multilinear map $\Phi:T^n(\Omega(X,\g))\to\Omega(X,\g)$ of 
order $n$ and degree $p$ is given by\footnote{At first sight the product 
does not seem to agree. However one has to understand $\star$ as 
acting in the middle. Pulling the (shifted) product out one has naturally 
$\tilde M(\tilde\omega_1,\tilde\omega_2)=(-1)^{p_1}\widetilde{\omega_1\star\omega_2}$.} 
\begin{equation}
  \Phi\oint(\omega_1,\cdots,\omega_m)=
  \sum_k(-1)^{(p+n-1)\eps_k}\oint(\omega_1,\cdots,\omega_k,\tilde\Phi(\omega_{k+1},\cdots,\omega_{k+n}),\omega_{k_{n+1}},\cdots,\omega_m). 
\end{equation}
The signs are a generalization of those appearing in the 
equation for $\nabla$ and in the wedge product. 
They suggest that we should assign a degree $p+n-1$ to $\Phi$. 
Note that for the operators appearing in the covariant derivative 
this degree is $p+n-1=1$.
We will see below that this degree is natural in a particular shifted 
version of the Hochschild complex. 

This is similar to the situation of a bundle on $X$. There the 
connection could be viewed as a map on $\Omega(X,\g)$ of 
degree 1. In that case we know that a connection can equivalently 
be characterized to be such a map satisfying the Leibniz rule. 

In this case we have a similar characterization. Let us formulate 
the Leibniz rule for this case. Any function $f$ on $X$, and more 
generally any differential form, induces a multiplication map 
which we can view as an element of $\End(\Omega(X,\g))$. 
Using the above this induces a multiplication operator $m_f$ 
on the tensor algebra $T(\Omega(X))$. In terms of this multiplication 
map, the covariant derivative satisfies the Leibniz rule 
\begin{equation}
  \nabla\circ m_f -m_f\circ\nabla = m_{df}.
\end{equation}
This means that if we write $\nabla=d+\CB$, then $\CB$ 
should commute with the multiplication. It therefore 
should be local, and becomes (locally) an element of 
\begin{equation}
  \CB\in\Omega(X)\otimes\End(T(\g),\g)=\Omega(X)\otimes C(\g,\g),
\end{equation}
having total degree 1. In other words, the connection form is 
a differential with values in the \emph{Hochschild complex} 
$C(\g,\g)$ of multilinear maps.

\subsection{Hochschild Complex}

We saw above that the connection form on loop space naturally is a 
differential form with values in the Hochschild complex. 
For any  associative algebra $\g$, the Hochschild complex is the space 
of multilinear maps on $\g$, $C^n(\g,\g)=\Hom(\g^{\otimes n},\g)$. This complex has a 
canonical twisted Lie bracket, called the Gerstenhaber bracket, extending the 
commutator in $C^1(\g,\g)=\End(\g)$. This bracket is denoted $[\cdot,\cdot]$, 
its definition can be found in \appref{app:hoch}.
Let us list a few special cases for this bracket, to get an idea of how it acts. 
On elements $\alpha\in C^0(\g,\g)=\g$ and $\mu\in C^1(\g,\g)=\End(\g)$ it 
gives the action $[\mu,\alpha] = \mu(\alpha)$. For 
$\phi\in C^2(\g,\g)=\Hom(\g^{\otimes 2},\g)$, $[\phi,\phi]$ is the associator 
of the product $\phi$. Furthermore $[\phi,\alpha]=\phi(\alpha,\cdot)-\phi(\cdot,\alpha)$ 
is the commutator of $\alpha$ with respect to the product $\phi$, and 
$[\mu,\phi]$ is the derivation of the product $\phi$ by the linear map $\mu$. 
The general definition of the Gerstenhaber bracket is given in the Appendix. 

When $\g$ is an associative algebra with product $M$, the Hochschild complex 
has a nontrivial coboundary operator $\delta=[\cdot,M]$ called the 
Hochschild differential. It can easily be seen that $\delta^2=0$ is 
equivalent to the associativity $[M,M]=0$ of the product. 

It will be natural to consider $\g$ to be a graded algebra, such as the 
algebra of differential forms. Above we saw at various places 
that in the relation to the loop space the (form) degree was naturally 
shifted by one. This was basically due to the integration over the loop, 
eating up one degree. It is therefore natural in this context to consider 
the shifted complex $\Pi\g$, which is defined by shifting the degree by 
one according to $(\Pi\g)^p=\g^{p+1}$. In fact this shift is natural also 
from the point of view of the Hochschild complex for the following reason. 
The Gerstenhaber bracket is not exactly a (graded) Lie bracket, 
as the signs are somewhat peculiar. However, if one goes to the 
shifted complex the Gerstenhaber bracket becomes a true graded Lie bracket. 
Any map $\phi\in C(\g,\g)$ is related to a map $\tilde\phi$ 
on the shifted space $\Pi \g$ through $\tilde\phi=\Pi\phi(\Pi\inv)^{\otimes n}$. 
The degree of the shifted map is given by $|\tilde\phi| = |\phi|+n-1$. 
There is a natural composition of shifted multilinear maps, which we 
denote by $\circ$, defined in \appref{app:hoch}. 
It is defined by viewing the multilinear maps 
$\tilde\phi$ as graded multi-derivations of the tensor algebra of $\Pi \g$. 
Using this composition and the natural grading, the super commutator 
defines a natural graded Lie-bracket, 
\begin{equation}
  \lb\tilde\phi_1,\tilde\phi_2\rb = 
  \tilde\phi_1\circ\tilde\phi_2-(-1)^{|\tilde\phi_1||\tilde\phi_2|}\tilde\phi_2\circ\tilde\phi_1.
\end{equation}
For $\phi_i\in C^{n_i}(\g,\g)$, the Gerstenhaber bracket can be related to 
the super commutator on the shifted complex by the relation 
\begin{equation}
  \widetilde{[\phi_1,\phi_2]} = (-1)^{(n_1-1)|\phi_2|}\lb\tilde\phi_1,\tilde\phi_2\rb.
\end{equation}

\section{Connections on Algebra Bundles}
\label{sec:algbundle}

In this section we will analyze the gauge theory that is defined by the 
connection based on differential forms with values in the Hochschild complex.

\subsection{Generalized Connections and Curvature}

In the last section we saw that, under certain locality assumptions, 
bundles on loop space $\CL X$ give rise to algebra bundles on the manifold $X$. 
The connection on this algebra bundle has a homogeneous part which 
is a differential form with values in the Hochschild complex 
$C(\g,\g)$ of the fiber $\g$. 

The connection on the algebra bundle corresponds to an element of this double 
complex of total degree 2. It can be decomposed into three components of 
bidegree $(2,0)$, $(1,1)$ and $(0,2)$ respectively, 
\begin{eqnarray}
  B \in \Omega^2(X,\g),\qquad
  A \in \Omega^1(X,\End(\g)),\qquad
  M \in \Omega^0(X,\Hom(\g^{\otimes2},\g)).
\end{eqnarray}
Here we included the product $M$ as a field. This allows us to describe nontrivial 
algebra bundles, such as Azuyama algebras. From this point of view this 
is a natural extension. Connections of precisely this form can also be 
found by making the global description of Gireauds nonabelian gebes in 
terms of transition functions infinitesimal \cite{kon}. 

In our discussion of loop space we found that the signs involved in 
the action of the covariant derivative are those of the shifted 
Hochschild complex, $C(\Pi\Omega(X,\g),\Pi\Omega(X,\g))$. It is therefore 
natural to consider the covariant derivative as an element of this shifted 
complex, and introduce the shifted connection 
\begin{equation}
  \nabla=d+\fB=d+\tilde B+\tilde A+\tilde M. 
\end{equation}
Note that the shifted connection forms and the De Rham differential all have 
homogeneous degree 1. The twisted complex $C(\Pi \g,\Pi \g)$ has a canonical 
Lie bracket $\lb\cdot,\cdot\rb$. This structure is indeed similar to Yang-Mills 
connections, and makes it easy to find curvatures, gauge transformations, 
and the Bianchi identities. 

Let us start with the curvature. The curvature should be defined as the 
square of the covariant derivative, $\fH=\nabla^2$. Here one 
should use the composition $\circ$ in the (twisted) Hochschild 
complex. As the connection forms have odd total degree they anticommute. 
Hence their product can be written in terms of the Lie bracket. 
This leads to an expression which looks very much like the one we are 
used to in Yang-Mills, namely 
$\fH=d\fB+\frac{1}{2}\lb\fB,\fB\rb$. 
In components and after twisting back we find 
\begin{eqnarray}
  \CH &=& dB+A(B),\nonumber\\
  \CF &=& dA+A^2+[M,B],\nonumber\\
  \CN &=& dM+[A,M],\\
  \CK &=& \frac{1}{2}[M,M].\nonumber
\end{eqnarray}
Here we used that $A^2=\frac{1}{2}[A,A]$ and $A(B)=[A,B]$. 
Note that signs are naturally introduced by the shifting defined 
above.  For example, half of the $A(B)$ term is actually of the form 
$-[B,A]$. If we would have simply written the quadratic term in the 
curvature in the untwisted complex, as $[B+A+M,B+A+M]$, the 
contributions would actually cancel instead of add up. 
This would not lead to a consistent nonabelian 2-form theory. 
We see that $A$ plays the role of connection, covariantizing the De Rham 
differential $d$ to $d_A=d+[A,\cdot]$. Also we see the emergence of the 
Hochschild differential $\delta_M=[M,\cdot]$ for the local product $M$. 
The last component $\CK$ can be identified with the associator of $M$. 

The structure of differential Lie algebra on the shifted complex 
ensures that the curvature satisfies a Bianchi identity. 
It has the familiar looking form $d\fH+\lb\fB,\fH\rb=0$. 
In its full generality, the homogeneous components of the Bianchi 
identity can be written in the untwisted complex as 
\begin{eqnarray}
  d\CH+[A,\CH]+[B,\CF] &=& 0,\nonumber\\
  d\CF+[A,\CF]-[B,\CN]-[M,\CH] &=& 0,\nonumber\\
  d\CN+[A,\CN]+[B,\CK]+[M,\CF] &=& 0,\\
  d\CK+[A,\CK]-[M,\CN] &=& 0,\nonumber\\{}
  [M,\CK] &=& 0.\nonumber
\end{eqnarray}

The various fields that appear in the gauge theory of algebra bundles 
are summarized in \tbref{tb:algfields}, indicating their form degree 
and the part of the Hochschild complex where they take values. 

\begin{table}[ht]
\[
\begin{array}{r|cccc}
 & \Omega^0(X) & \Omega^1(X) & \Omega^2(X) & \Omega^3(X) \\ \hline
\g & k & \alpha & B & \CH \\
\End(\g) & \gamma & A & \CF & \\
\Hom(\g^{\otimes 2},\g) & M & \CN &  & \\
\Hom(\g^{\otimes 3},\g) &\CK  &  &  & \\
\end{array}
\]
\caption{The fields in the gauge theory for algebra bundles.}
\label{tb:algfields}
\end{table}

\subsection{Gauge Transformations}

Writing down the gauge transformations is just as easy. The gauge 
transformations are generated by elements of total degree 1, 
or degree zero in the shifted complex. 
This has the two homogeneous components 
\begin{equation}
  \alpha \in \Omega^1(X,\g),\qquad
  \gamma \in \Omega^0(X,\GL(\g)).
\end{equation}
The shifted elements generating infinitesimal gauge transformations 
will be combined in the form $\fA=\tilde\alpha+\tilde\gamma$. 
The infinitesimal gauge transformations can then be given as the 
covariant derivative of the generator $\delta\fB=d\fA+\lb\fB,\fA\rb$. 
This can of course be easily recast in terms of the unshifted 
variables. Integrating these infinitesimal gauge transformations 
we find the following finite gauge transformations for the components 
of the gauge connection 
\begin{eqnarray}
  B' &=& \gamma\inv(B+d\alpha+A(\alpha)-M(\alpha,\alpha)),\nonumber\\
  A'  &=& \gamma\inv(d\gamma+A\gamma-[M,\alpha]\gamma),\\
  M'  &=& \gamma\inv M\gamma^{\otimes 2}.\nonumber
\end{eqnarray}
Note that $A(\alpha)=[A,\alpha]$ and $M(\alpha,\alpha)=\frac{1}{2}[[M,\alpha],\alpha]$. 
The finite gauge transformations generated by the 1-forms is 
given by the exponential $\exp(-[\alpha,\cdot])$. 

Let us denote a gauge transformation with generators $(\gamma,\alpha)$ by $U_{\gamma,\alpha}$. 
Note that $U_{\gamma,\alpha}=U_{\gamma,0}U_{1,\alpha}$. Furthermore, the gauge 
transformations satisfy $U_{\gamma,0}U_{1,\alpha}=U_{1,\gamma\inv\alpha}U_{\gamma,0}$. 
Hence the gauge group is the semi-direct product of 
the group of local $\GL(\g)$ transformations generated by $U_{\gamma,0}$ 
and the abelian ``translation'' group generated by $\alpha$,  
\begin{equation}
  \Omega^0(X,\GL(\g))\ltimes\Omega^1(X,\g).
\end{equation}
This also shows that the gauge transformations close among themselves. 
(The reader should beware that $U$ as defined here is an antihomomorphism). 

The differential Lie structure guarantees that the infinitesimal gauge 
transformation of the curvature is $\delta\fH=\lb\fH,\fA\rb$. 
Hence the transformation of the curvature is strictly locally, 
that is without derivatives. The untwisted components of the curvature 
have finite gauge transformations of the form 
\begin{eqnarray}
  \CH' &=& \gamma\inv\Bigl(\CH+\CF(\alpha)-\CN(\alpha,\alpha)-\CK(\alpha,\alpha,\alpha)\Bigr),\nonumber\\
  \CF' &=& \gamma\inv\Bigl(\CF-[\CN,\alpha]-\frac{1}{2}[[\CK,\alpha],\alpha]\Bigr)\gamma,\nonumber\\
  \CN' &=& \gamma\inv(\CN+[\CK,\alpha])\gamma^{\otimes 2},\\
  \CK' &=& \gamma\inv\CK\gamma^{\otimes 3}.\nonumber
\end{eqnarray} 
Also note that $\CN(\alpha,\alpha)=\frac{1}{2}[[\CN,\alpha],\alpha]$ 
and $\CK(\alpha,\alpha,\alpha)=\frac{1}{6}[[[\CK,\alpha],\alpha]\alpha]$.

As we are dealing with a 2-form gauge theory, the gauge transformations above 
are not all independent. There will be gauge-for-gauge transformations 
acting on the gauge parameters $(\gamma,\alpha)$. They are generated by 
functions 
\begin{equation}
  k \in \Omega^0(X,\g). 
\end{equation}
Using the twisted form $\fC=\delta\tilde k$ we can write the infinitesimal 
transformations of the gauge parameters in the form $\delta\fA=d\fC+\lb\fB,\fC\rb$. 
On the untwisted fields this becomes 
\begin{eqnarray}
  \delta\alpha &=& dk + A(k) + [[M,k],\alpha],\nonumber\\
  \delta\gamma &=& [M,k]\gamma.
\end{eqnarray}

These gauge-for-gauge transformations, in contradistinction to the abelian case, 
do not leave the connection invariant. To see this, we look at the effect of 
the combination of a deformed gauge transformation $U_{\alpha',\gamma'}$, 
where $\alpha'=\alpha+\delta\alpha$ and $\gamma'=\gamma+\delta\gamma$, 
and the inverse of the undeformed gauge transformation 
$U_{\alpha,\gamma}\inv=U_{-\gamma\inv\alpha,\gamma\inv}$. We have 
\begin{equation}
  U_{\alpha,\gamma}\inv U_{\alpha',\gamma'} = U_{\alpha'',\gamma''};\qquad
  \gamma'' = 1+[M,k],\quad \alpha'' = dk+A(k).
\end{equation}
With respect to these transformations, the gauge connections 
transform as 
\begin{equation}
  \delta B = \CF(k),\qquad
  \delta A = 0,\qquad
  \delta M = 0.
\end{equation}
There are similarly transformations for the components of the field strength. 
Note that the action of the gauge-for-gauge transformations on the connection 
fields and the curvatures vanishes when we would have $\CF=0$. 

The finite gauge-for-gauge transformations are not strightforward to 
write down in general. For this we need to be able to exponentiate 
the product with $k$ using $M$. In a more restricted situation we will 
give the finite transformation later.

\section{Reductions}
\label{sec:crosmod}

In the above we saw that the gauge transformations $\gamma$ were based on a gauge 
group $\GL(\g)$ acting on a vector space $\g$. We can make generalizations of this 
based on reduced gauge groups, or more precisely a reduction of the full 
Hochschild complex.

\subsection{Reductions and Crossed Modules}

We will now take $M$ to be a constant product. 
Therefore $\g$ will be a fixed algebra. Note that in the formulas we wrote 
down the product appears only in a commutator. It will therefore be 
more natural to use the Hochschild differential $\delta=[\cdot,M]$ induced by the 
product. Indeed, the action of the components of the connection form 
act via the bracket. The covariant derivative can be consistently replaced by 
\begin{equation}
  \nabla = d+\delta+\fB+\fA.
\end{equation}
For example, the formula for the 2-form curvature will be written 
$\CF = dA+A^2+\delta(B)$, which is easily seen to correspond to the 
2-form part of $\nabla^2$ with $\nabla$ as above. It is important 
that $\delta$ is a derivation of the Lie bracket. In fact we have replaced 
the Hochschild complex with zero differential by the more general Hochschild 
complex with differential $\delta$. 

In the case of vector bundles, gauge bundles with gauge group $G$ 
can be constructed as special cases of $\GL$ bundles by 
reducing the holonomy to a subgroup. We want to do the same here, 
reducing the gauge group $\GL(\g)$ (whose Lie algebra is $\End(\g)$) 
to some smaller group $H$. That is, we reduce the 1-form connection 
$A$ to lie in $\Omega^1(X,\h)$. 
The internal vector space $\g$ where $B$ takes its values will be some 
representation space of $H$. 

We saw that the internal space of the connection forms 
is the complex $C^*(\g,\g)$. In general we should not just reduce the 
degree 1 part $\End(\g)$ of the complex, but the whole complex. 
Removing the product $M$ as a field reduces it to the first two terms. 
So the actual internal space $C^*(\g,\g)$ is reduced to a 2-term complex 
$\g\stackrel\delta\to\h$. For this reduction to be consistent there are several 
conditions. As remarked before, what made our setup work was the fact 
that we had the structure of a differential Lie algebra on the internal space. 
This property should be preserved by the reduction. 
The essential constraint on the reduced data therefore is that the 
2-term complex is supplied with a Lie bracket of the 
proper degree such that the coboundary $\delta$ is a derivation.

Let us analyze these constraints. As the Lie bracket 
has degree $-1$, one of its arguments must be in $\h$; 
therefore there is only room for two components. 
The restriction $\h\times \h\to \h$ provides a Lie 
bracket on $\h$. The mixed bracket $\h\times \g\to \g$ 
defines a left-action of $\h$ on $\g$. It follows from 
the Jacobi identity of the total bracket that this 
action is a representation of the Lie algebra $\h$. 
The map $\delta$ should be a (twisted) derivation 
of this bracket. This gives the two conditions 
\begin{equation}
  [\delta(X_1),X_2] = -[X_1,\delta(X_2)],\qquad 
  \delta([Y,X_2]) = [Y,\delta(X_2)],
\end{equation}
for $X_1,X_2\in\g$ and $Y\in\h$. The first identity shows that there is an 
induced antisymmetric bracket on $\g$ defined by 
\begin{equation}
  [X_1,X_2]_\g = [\delta(X_1),X_2],\qquad X_1,X_2\in\g. 
\end{equation}
Using the Jacobi identity for $[\cdot,\cdot]$ and the second relation, 
we find that $[\cdot,\cdot]_\g$ satisfies the Jacobi identity. 
Hence also $\g$ is a Lie algebra. Taking $Y=\delta(X_1)$, 
the second relation implies that $\delta$ is a homomorphism of Lie algebras, 
\begin{equation}
  [\delta(X_1),\delta(X_2)] = \delta([\delta(X_1),X_2])=\delta([X_1,X_2]_\g).
\end{equation}

The above description should be understood only as infinitesimal 
relations. We can integrate the Lie algebras $\g$ and $\h$ 
to groups $G$ and $H$. Then $\delta$ will induce 
a group homomorphism 
\begin{equation}
  G\stackrel{\delta}{\rightarrow} H.
\end{equation}
The action of $\h$ on $\g$ will integrate to a left-action of $H$ on 
the group  $G$, which we will denote ${}^hg$ for $h\in H$ and $g\in G$. 
The identities above imply the integrated identities 
\begin{equation}
  {}^{\delta(g_1)}g_2 = g_1g_2g_1\inv,\qquad
  \delta({}^{h}g) = h\delta(g)h\inv. 
\end{equation}
The data above defines the notion of a \emph{crossed module}. 

There are several possibilities for a choice of crossed module. 
Let us fix the group $G$, which might be any group. 
We saw that $H$ acts on $G$ by automorphisms. Therefore, a canonical 
choice is $H=\Aut(G)$. The differential $\delta$ is given by inner 
conjugation: for $g\in G$ $\delta(g)$ is the automorphism conjugation by $g$. 
One easily checks that all the conditions of a crossed module are met 
for this case. A closely related crossed module is given by taking 
for $H$ just the inner automorphisms. A somewhat trivial crossed module 
is given by taking $H=G$. The action of $H$ on $G$ is now the conjugation 
automorphism. Any crossed module has a natural homomorphism to the first 
crossed module $G\to\Aut(G)$. Indeed, the left action of $H$ 
on $G$ defines a homomorphism to $\Aut(G)$, and the above two relations 
ensure this to define a homomorphism of crossed modules. So in a sense 
the first example is universal. Therefore we will mainly concentrate 
in this case. 

Let us shortly comment on how to implement this in the loop space picture 
that formed our motivation earlier. Note that we need the $\omega_i$ 
to take values in the Lie algebra $\g$. This is a Lie algebra, rather 
than an associatove algebra, which was needed to define the nonabelian 
product $\star$. The natural thing is to take the product in the 
universal enveloping algebra (UEA) $U(\g)$. Writing $\omega_i^W=\omega_i^{a_i}t_{a_i}$, 
where the $t_a$ are generators of the Lie algebra $\g$, we have  
\begin{equation}
  \oint_A(\omega_1,\cdots,\omega_n) 
  = \oint(\omega_1^{a_1},\cdots,\omega_n^{a_n})t_{a_1}\otimes\cdots\otimes t_{a_n},
\end{equation}
where the last expression is understood in the UEA $U(\g)$. 
Using these, the $\g$-valued 2-form $B$ acts naturally on 
these elements, as does the $\h$-valued $A$. 
Note that for the abelian situation using the UEA $U(\u(1))=S^*\R$  
would have been equivalent, as $S^n\R\cong\R$.

\subsection{Connections on Non-Abelian Gerbes}

Let us now see what the formul\ae{} we wrote down look like 
in the reduced situation. The various fileds appearing in the 
reduced gauge theory are summarized in \tbref{tb:gerfields}. 
The two curvature components can be written 
\begin{equation}
  \CH = dB + A(B),\qquad
  \CF = dA + A^2 + \delta(B).
\end{equation}
The Bianchi identities for these curvatures are 
\begin{equation}
  d\CH + A(\CH) - \CF(B) = 0,\qquad
  d\CF + [A,\CF] + \delta(\CH) = 0.
\end{equation}
The gauge transformations of the connection and curvature reduce to 
\begin{equation}
\renewcommand{\arraystretch}{1.2}
\begin{array}{r@{\;}c@{\;}l@{\qquad}r@{\;}c@{\;}l}
  B' &=& \gamma\inv(B + d\alpha + A(\alpha) - \alpha^2),&
  \CH' &=& \gamma\inv(\CH+\CF(\alpha)),\\
  A' &=& \gamma\inv(d\gamma + A\gamma - \delta(\alpha)\gamma),&
  \CF' &=& \gamma\inv\CF\gamma.
\end{array}
\end{equation}
For this situation it is rather straightforward to integrate 
the gauge-for-gauge transformation. They are given by 
\begin{equation}
  \alpha' = (dk + A(k) + k\alpha)k\inv,\qquad
  \gamma' = \delta(k)\gamma.
\end{equation}
These are equivalent to at least part of the equations for connections 
on nonabelian gerbes as found in \cite{bremes}. In that paper some 
more gauge symmetries were added, and conventions and notation 
were slightly different. 

\begin{table}[ht]
\[
\begin{array}{r|cccc}
 & \Omega^0(X) & \Omega^1(X) & \Omega^2(X) & \Omega^3(X) \\ \hline
\g & k & \alpha & B & \CH \\
\h & \gamma & A & \CF & \\
\end{array}
\]
\caption{The various fields in the reduced gauge theory.}
\label{tb:gerfields}
\end{table}

\section{BV Sigma Models}
\label{sec:bv}

In this section we will construct BV models for the 2-form gauge theories. 
The formulas above lead to canonical BV sigma models. After gauge fixing 
this can gives rise to actions of Yang-Mills, Chern-Simons, or BF type.

\subsection{BV Models for Algebra Bundles}

Let us now consider actions for the gauge theories we described in this paper. 
Having gauge symmetries, we will need a gauge fixing mechanism. Therefore 
we will need to add ghost and antighost fields to the theory. Gauge fixing 
is rather straightforward to do when one has a BV symmetry. Often in 
gauge theories, a BV actions is somehow constructed out of the classical action. 
Here we take a different road and start by constructing a BV action. 

The fields in our gauge theory are differential forms of various degrees. 
The gauge symmetries are generated by lower degree forms. More generally, 
any $p$-form gauge field will generate ghosts and corresponding antifields 
of various degrees. It will be convenient to summarize all these fields using 
superfields. We introduce supercoordinates $(x^\mu|\theta^\mu)$ 
of degree 0 and $-1$. Contracting the form indices of the 
fields we introduce superfields with an expansion of the form 
\begin{eqnarray}
  \Bphi = \phi_{(0)} + \theta\phi_{(1)} + \cdots + \theta^d\phi_{(d)},
\end{eqnarray}
where the component $\phi_{(p)}$ is a $p$-form and we suppressed contractions 
of indices. If the superfield $\Bphi$ has ghost degree $g$, then the ghost degree 
of $\phi_{(p)}$ is $g-p$. Note that this implies that the ``physical'' field 
of ghost number zero will have form degree $g$. Next we have to write down 
an action for them. In fact using the structure of the theory, the differential 
Lie algebra, makes it is rather straightforward to write down a BV model. 

For the most general model we had the three connection components 
components $B_{(2)}$, $A_{(1)}$, and $M_{(0)}$. Corresponding to 
each of them we introduce superfields, $\BB$, $\BA$ and $\BM$,
taking values in the corresponding spaces in $C^*(\g,\g)$. 
They have ghost degree 2, 1, and 0, respectively. To get a BV structure 
we introduce superfields containing the antifields, $\BU$, $\BV$, $\BW$, 
and $\BX$. They have ghost degree $d-3$, $d-2$, $d-1$, and $d$, 
respectively. More precisely, the antifield of a particular component 
is the component of the complementary degree---for example the 
antifield of $B_{(p)}$ is the component $U_{(d-p)}$. 
Note that the BV structure is degenerate, as $\BX$ is not paired up 
with any other superfield. In principle this is not a problem. One can 
for example introduce an auxiliary superfield of degree $-1$ containing 
its antifield, not appearing in the action. 

In order to write down an action, we need an inner product $\langle\cdot,\cdot\rangle$ 
on the Hochschild complex $C^*(\g,\g)$. This could for example be constructed 
by starting from some inner product on $\g$, and extending it to 
$C^n(\g,\g)\cong \g^{\otimes n}\otimes \g^*$. In fact at this stage, we 
could also have taken the antifields taking values in the dual space 
$C(\g,\g)^*=C(\g^*,\g^*)$. 

The BV action functional is constructed as to reflect the equations for the 
field strength and gauge transformations of the theory. The terms in the action 
of purely ghost number zero terms will be such that the variation of the 
ghost number zero components of the anti-superfields, $U_{(d-3)}$, 
$V_{(d-2)}$, $W_{(d-1)}$, and $X_{(d)}$, give the various components of the 
field strength. By replacing the ghost number zero fields by the full superfield, 
we find the complete BV action functional 
\begin{eqnarray}
  S_{BV} = \dint[_\CX]{d^d\!xd^d\!\theta} \Bigl( &&
    \langle\BU,\Bd\BB+\BA(\BB)\rangle
    + \langle\BV,\Bd\BA +\BA^2+[\BM,\BB]\rangle \nonumber\\
&&    + \langle\BW,\Bd\BM +[\BA,\BM]\rangle
    + \frac{1}{2}\langle\BX,[\BM,\BM]\rangle \Bigr),
\end{eqnarray}
where $\Bd=\theta^\mu\del_\mu$ is the De Rham differential 
and the integration is over the full superspace 
$\CX=\Pi TX$, so over both $x$ and $\theta$. This action functional has ghost 
degree zero and satisfies the (quantum) master equation 
$(S,S)+2i\hbar\Delta S=0$. This is a consequence of the Jacobi 
identity for the Gerstenhaber bracket, and the proof is equivalent to that 
of the Bianchi identity. 

The (linearized) gauge transformations can be recovered from the action of 
the BV BRST operator $\BQ=(S_{BV},\cdot)$ on the fields, that is 
$\delta\phi=(-1)^{(d-1)p}\frac{\del S_{BV}}{\del\phi^+}$. 
Here of course the gauge generators are replaced by the ghosts of ghost number one, 
$\alpha$ by $B_{(1)}$ and $\gamma$ by $A_{(0)}$. 
The higher order gauge transformations can be found from the exponentiation 
of the corresponding operation. We conclude that indeed the BV action given 
above exactly reflects the equations for the gauge theory. 

In order to get a physical theory we need to gauge fix the BV action. 
The gauge fixing is determined by a choice of fields and antifields 
and a gauge fixing fermion to fix the antifields. In fact one can get 
different physical theories from the above BV action functional, 
depending on the choice of fields. We will next discuss different 
choices one can make, and what kind of physical theories they lead to. 

Let us first focus on the field strength $\CH$. We will demonstrate 
how to find a Yang-Mills type action for this field strength from 
the above BV action. To get such an action we will take all 
components of $\BB$ to be fields and consider the components of $\BU$ 
as antifields. To get a Yang-Mills type action we need a gauge fixing 
that replaces $U_{(d-3)}=*\CH$. It is easy to write down a gauge fixing 
fermion that accomplishes this; it contains the term 
\begin{equation}
  \Psi = \int_\CX \langle B_{(d-3)} ,*(dB+A(B))\rangle + \cdots .
\end{equation}
This implies that after gauge fixing the antifield is replaced by 
\begin{equation}
 U_{(d-3)} = *(dB+A(B)) = *\CH. 
\end{equation}
Therefore, the gauge fixed action will contain the following 
term involving only physical fields, 
\begin{equation}
  S_{gf} = \int_X \Bigl( \langle\CH,*\CH\rangle + \cdots \Bigr). 
\end{equation}
Hence we have introduced a Yang-Mills type of quadratic 
kinetic term for the 3-form field strength. 

Another procedure is to regard $B_{(d-3)}$ as an antifield and 
$U_{(d-3)}$ correspondingly as a field. More generally, we 
take all fields with negative ghost number in $\BB$ and $\BU$ 
to be antifields. This will lead to a topological theory. 
As $U_{(d-3)}$ is a field, the gauge fixed action will contain 
the term 
\begin{equation}
  S_{gf} = \int_X \Bigl( \langle U_{(d-3)},\CH\rangle + \cdots \Bigr). 
\end{equation}
If we choose the gauge fermion $\Psi$ independent of the field $U_{(d-3)}$, 
this will be the only term in the gauge fixed action depending on it. 
We find a BF type theory. We can regard the field $U_{(d-3)}$ as a 
Lagrange multiplier, constraining field strength $\CH$ to zero. 
So this is a way to get to a flat gauge field. 

Analogous constructions apply to the other field strengths. 
Depending on whether we choose either of 
$V_{(d-2)}$, $W_{(d-1)}$, and $X_{(d)}$ as field or antifields, 
we can generate either Yang-Mills type terms in the action for 
the field strengths $\CF$, $\CN$ and $\CK$, or Lagrange multiplier terms. 

The fact that we can write down a BV action functional, 
satisfying the BV quantum master equation, which gives after 
gauge fixing a $\CH*\CH$ term, shows that in fact we have 
a well-defined quantum gauge theory involving the 2-form. 
This is a very nontrivial statement indeed, and seems to be 
the only nontrivially interacting quantum gauge theory 
for 2-forms gauge fields. 

Of course in gauge fixing we should also take care of the gauge invariance 
of the forms. For this we need to add extra antighost and Lagrange multiplier 
terms to the action, and gauge fixing terms in the gauge fixing fermion. 
This will be rather straightforward and standard, so we will not explicitly 
write this part out. But let us illustrate the procedure for the gauge field $B_{(2)}$. 
We introduce an antighost $\overline b_{(1)}$ and Lagrange multiplier term 
$\underline b_{(d-1)}$, along with the antifields $\overline b_{(d-1)}^+$ and 
$\underline b_{(1)}^+$. We then add an antighost term to action of the form 
$S_{ag}=\int \langle \underline b_{(1)},\overline b_{(d-1)}^+\rangle$, 
and include the following term in the gauge fermion, 
\begin{equation}
  \Psi = (-1)^{d}\int \langle d\overline b_{(1)},*B_{(2)}\rangle+\cdots.
\end{equation}
The gauge fixing will then set $\overline b_{(d-1)}^+= d*B_{(2)}$. 
Substituting this in the antighost term of the action gives the gauge fixing 
term for the Lorentz-type gauge in the action.

\subsection{BV Models for Gerbe Connections}

We can now specialize to the reduced case 
of a nonabelian gerbe. The BV action still goes through, 
with only the superfields $\BB$, $\BA$, $\BU$, and $\BV$, 
with $\BB$ and $\BH$ taking values in $\g$ and $\BA$ and 
$\BV$ taking values in $\h$. 
Also for the gauge fixing we draw the same conclusion; 
we can either get a quadratic Lagrangian, or constrain 
the field strength to zero. 

To construct the action we need a trace, or rather an inner product, 
on the crossed module. This means we need inner products on 
$\g$ and $\h$. These inner products can be chosen to be invariant under 
the action of $\h$ by the bracket, at least when $H$ is compact. 

The BV action above reduces to 
\begin{equation}
  S_{BV} = \dint[_\CX]{d^d\!xd^d\!\theta} \Bigl( 
   \langle\BU, \Bd\BB+\BA(\BB)\rangle_\g + \langle\BV, \Bd\BA+\BA^2+\delta(\BB)\rangle_\h \Bigr). 
\end{equation}

As above we can get several different physical theories depending 
on the gauge fixing procedure, resulting in Yang-Mills type actions or 
Lagrange multipliers for the field strengths $\CH$ and $\CF$. 
Let us illustrate this for a particular choice. 
Let us consider all fields in $\BU$ as antifields, and all fields of 
negative ghost number in $\BV$ and $\BA$ as antifields. 
We choose the gauge fixing fermion 
\begin{equation}
  \Psi = \int_X \Bigl(\langle B_{(d-3)},*(dB_{(2)}+A_{(1)}(B_{(2)}))\rangle_\g \Bigr).
\end{equation}
The antifield $U_{(d-3)}$ is fixed to $U_{(d-3)}=*\CH$, 
where $\CH$ is given in terms of $B_{(2)}$ and $A_{(1)}$ 
as before. The gauge fixed action has the form 
\begin{equation}
  S_{gf} = \int_X \biggl(\langle\CH,*\CH\rangle_\g + \langle V_{(d-2)},\CF\rangle_\h +\cdots \biggr),
\end{equation}
where the dots denote terms containing ghosts and antighosts. 
These are precisely the kinetic term for $B_{(2)}$ and the 
Lagrange multiplier term for $\CF$. The most natural dimension 
for this is $d=6$, as then all the terms have precisely dimension 
6 and therefore are marginal. In dimension less than 6, 
the interactions are irrelevant, while in higher dimensions 
the interactions are all relevant. 

In $d=5$ we also have another possibility. Let us consider 
$B_{(3)}$ as an antifield, and take the gauge fermion to be 
\begin{equation}
  \Psi = \int_X \Tr\Bigl( \langle B_{(3)},B_{(2)}\rangle_\g +\cdots \Bigr).
\end{equation}
We then find $U_{(2)}=B_{(2)}$, and the gauge fixed action 
contains the field terms 
\begin{equation}
  S_{gf} = \int_X \Tr\biggl( \langle B_{(2)},dB_{(2)}\rangle_\g 
  + \langle B_{(2)},A_{(1)}(B_{(2)})\rangle_\g + \langle V_{(d-2)},\CF\rangle_\h +\cdots \biggr).
\end{equation}
This is a 2-form Chern-Simons theory for $B_{(2)}$. 

\section{Concluding Remarks}

In this paper we constructed nonabelian 2-form connections. 
The main emphasis in this paper was on some basic formulas of 
the curvature and gauge transformations. These 
formulas we derived were extracted from bundles on loop space. 

As mentioned in the introduction we expect nonabelian 2-forms 
to play a role in string theory for stacks of 5-branes. 
It would be interesting to see whether the theories constructed 
in this paper can indeed be shown to describe the physics 
of these systems in some decoupling limit. 
The $(2,0)$ LST describing NS5A- and M5-brane systems 
are strongly coupled, and we do not expect a local field theory 
of 2-forms to describe the local physics in these situations. 
It has however been argued that the IR physics should be described 
by 2-forms. Therefore we should expect that at least on the level 
of topology nonabelian 2-form connections are important. In other words, 
the topology, such as characteristic classes and holonomy, could 
describe nontrivial configurations of 5-branes. We expect therefore 
that for such questions our theory is relevant to 5-brane systems. 

We saw that for the construction of a theory based on a particular 
gauge group $G$, we needed extra data. The connection is 
described by a crossed module $G\stackrel\delta\to H$ rather than just $G$. 
Therefore one should ask the question which crossed module 
is actually relevant for decoupled 5-brane systems in string theory. 
The most natural one is the crossed module $G\to\Aut(G)$. 
It is universal in the sense that any other crossed module has 
a homomorphism to one of this form. This would therefore be the 
most natural guess. The choice of $H$ certainly has an effect 
on the topology of 2-form connections. For example when the 
manifold has a nontrivial fundamental group there can be 
nontrivial monodromies with values in $H$. 
One can indeed argue based on the presence 
of such classes \cite{hof2} that for the double scaled 
little string theories \cite{givkut} the crossed module 
$G\to\Aut(G)$ gives the correct topology of wrapped 5-branes. 
For $H=\Aut(G)$ the main effect comes from the outer automorphisms, 
which leads to certain discrete cohomology classes. 
For a mathematical construction of the nonabelian cohomology involved 
and related to global nonabelian gerbes, see for example \cite{gir,breen}. 

We found several different gauge fixed actions for the gauge theories. 
It would be interesting to see whether any of them can describe 
decoupling limits of 5-brane theories. It is encouraging that 
one can find Yang-Mills type theories, as these seem to be relevant 
at least in the weak field limit of abelian 2-forms. We should also 
mention that none of the (partially) gauge fixed actions preserve 
the gauge transformations generated by $\alpha$---only the subgroup 
generated by abelian $\alpha$ is preserved. The nonabelian gauge 
transformations are however broken by the gauge fixing procedure, 
this therefore is not necesssarily a problem. It suggests that the nonabelian 
1-form gauge symmetries will not be manifest in a physial field theory, 
though it plays an important role in organizing the theory. 

The discussion in this paper was merely local. In order to describe 
the topology of nonabelian 2-forms we need a global description. 
This can most easily be given in tems of a \v{C}hech description. 
Transition functions are easily found starting from the gauge 
transformations of the theory, as will be worked out in a later 
paper \cite{hof2}. Another aspect of global 
quantities are the ``nonabelian Wilson surfaces'', integration of 
the connection over a surface, giving some element of the group $G$. 
As surfaces do not have a natural ordering this is not easily achieved, 
and one needs either some extra restrictions or equivalences 
\cite{hof2}. Such questions were also recently discussed 
in \cite{chep,attal}.

\section*{Acknowledgements}

I am happy to thank Iouri Chepelev, Bill Messing, Larry Breen, 
Jae-Suk Park and Koen Schalm for enlightening 
discussions. This research was supported in part by DOE grant \#DE-FG02-96ER40959.

\appendix

\section{Hochschild Complex, Gradings and Signs}
\label{app:hoch}

Here we derive the signs in the bracket in the Hochschild complex 
$C^*(A,A)$ of a graded algebra $A$.

For an element $\alpha\in A$, we denote by $|\alpha|$ its degree. 
First, we use the shift operator $\Pi$ to identify the graded vector space $A$ 
with the shifted graded vector space $\Pi A=A[1]$. On the tensor product, 
this map is given on a homogeneous element of the tensor algebra of $A$ by 
\begin{equation}\label{shiftalg}
  \Pi^{\otimes n}(\alpha_1\otimes\cdots\otimes\alpha_n) = 
  (-1)^{\sum_k(n-k)|\alpha_k|}\Pi\alpha_1\otimes\cdots\otimes\Pi\alpha_n
\end{equation}
the signs are coming from commuting $\Pi$ through the $\alpha_k$. 
This map is such that it maps an element of the antisymmetrized 
tensor algebra $\ext[*]A$ to the symmetrized shifted tensor algebra 
$S^* A[1]$. In the following we will denote $\Pi\alpha\equiv\tilde\alpha$. 
Note that $|\tilde\alpha|=|\alpha|-1$, hence $\Pi$ has degree $-1$. 

This also induces a map between the multilinear operations. For any $n$-ary 
map $\phi$ on $A$ we have a naturally associated $n$-ary map $\tilde\phi$ on the 
shifted algebra $\Pi A$ defined by 
\begin{equation}\label{shiftmap}
  \tilde\phi = \Pi\circ\phi\circ(\Pi\inv)^{\otimes n}. 
\end{equation}
The gradings are related as $|\tilde\phi|=|\phi|+n-1$. In the shifted 
algebra, we can naturally define a composition of multilinear maps 
by extending them to graded multi-derivations on the tensor product 
$\bigoplus_n(\Pi A)^{\otimes n}$. 

Let us work out the composition in general. This will also explain the formula 
for the composition in $C^n(A,A)$. Note that the composition of two elements 
$\phi_i\in C^{n_i}(A,A)$, $i=1,2$, is of order $N:=n_1+n_2-1$. 
The graded derivation property of $\tilde\phi_2$ implies 
\begin{eqnarray}\label{compshift}
\tilde\phi_1\circ\tilde\phi_2(\tilde\alpha_1,\ldots,\tilde\alpha_{N}) 
&=& \sum_k(-1)^{\sum_{i=1}^k|\tilde\phi_2||\tilde\alpha_i|} 
 \tilde\phi_1(\tilde\alpha_1,\ldots,\tilde\alpha_i,
   \tilde\phi_2(\tilde\alpha_{k+1},\ldots,\tilde\alpha_{k+n_2}),
   \tilde\alpha_{n_2+k+1},\ldots,\tilde\alpha_{N}) 
\nonumber\\
&=& \sum_k(-1)^{\epsilon_k}
 \Pi\Bigl(\phi_1(\alpha_1,\ldots,\alpha_k,
   \phi_2(\alpha_{k+1},\ldots,\alpha_{k +n_2}),
   \alpha_{n_2+k+1},\ldots,\alpha_{N})\Bigr),
\end{eqnarray}
where the sign can be found by using \eqref{shiftalg} and \eqref{shiftmap}, 
and is given by 
\begin{equation}
\epsilon_k = \sum_{i=1}^{N}(N-i)|\alpha_i| + 
  |\phi_2|(n_1-1)+\sum_{i=1}^k|\phi_2||\alpha_i|-(n_2-1)k. 
\end{equation}
The first sum is the usual one coming from the commutation 
of the $\Pi$'s as explained above. 
Only the last two terms depend on the position $k$ of the $\phi_2$ in the chain. 
These therefore have to be contributed to the action of $\phi_2$ on the 
general element of the tensor algebra. Note that this has the usual term 
coming from commuting graded algebras, and an extra term depending on the order $n$. 
This motivates to define the composition of two elements in $C^*(A.A)$ as 
\begin{eqnarray}\label{comp}
&& \phi_1\circ\phi_2(\alpha_1,\ldots,\alpha_N) = \\
&&\qquad \sum_k(-1)^{(n_2-1)k+\sum_{i=1}^k|\phi_2||\alpha_i|}
 \phi_1(\alpha_1,\ldots,\alpha_k,\phi_2(\alpha_{k+1},\ldots,\alpha_{k+n_2}),\alpha_{k+n_2+1},\ldots,\alpha_N). \nonumber
\end{eqnarray}
This is natural, as it starts with a plus sign for $i=0$. 
Using this definition the relation \eqref{compshift} can be written 
\begin{equation}
 (-1)^{\sum_{i=1}^N(N-i)|\alpha_i|}
 \tilde\phi_1\circ\tilde\phi_2(\tilde\alpha_1,\ldots,\tilde\alpha_{N}) 
 = (-1)^{(n_1-1)|\phi_2|}\Pi\Bigl(\phi_1\circ\phi_2(\alpha_1,\ldots,\alpha_{N})\Bigr).
\end{equation}
This can also be written as 
\begin{equation}
 \phi_1\circ\phi_2
  = (-1)^{(n_1-1)|\phi_2|}\Pi\inv\circ\tilde\phi_1\circ\tilde\phi_2\circ\Pi^{\otimes(n_1+n_2-1)}.
\end{equation}
Notice the extra sign in the RHS. 

The super commutator of the composition provides the shifted Hochschild complex 
of the $\tilde\phi$ with a natural graded Lie-bracket, 
\begin{equation}
  \lb\tilde\phi_1,\tilde\phi_2\rb = 
  \tilde\phi_1\circ\tilde\phi_2 -(-1)^{|\tilde\phi_1||\tilde\phi_2|}\tilde\phi_2\circ\tilde\phi_1. 
\end{equation}

The Gerstenhaber bracket can now be defined using a similar relation
to this graded Lie-bracket (supercommutator) on the shifted algebra. 
We will also explicitly introduce the same sign factor we found above 
in the relation between the composition, 
\begin{equation}
  \widetilde{[\phi_1,\phi_2]} = (-1)^{(n_1-1)|\phi_2|}\lb\tilde\phi_1,\tilde\phi_2\rb 
  = (-1)^{(n_1-1)|\phi_2|} \Bigl(\tilde\phi_1\circ\tilde\phi_2 
  -(-1)^{|\tilde\phi_1||\tilde\phi_2|}\tilde\phi_2\circ\tilde\phi_1\Bigr).
\end{equation}
Using the relation between the compositions in the two pictures we find 
\begin{equation}
  [\phi_1,\phi_2] = \phi_1\circ\phi_2 -(-1)^{(n_1-1)(n_2-1)+|\phi_1||\phi_2|}\phi_2\circ\phi_1 
\end{equation}
This can be interpreted as a double graded supercommutators, with the degrees 
$(n-1,|\phi|)$. These are indeed the natural gradings in the Hochschild complex.

\end{document}